\documentstyle[aps,preprint,psfig]{revtex}
\newcommand{\be}{\begin{equation}}
\newcommand{\ee}{\end{equation}}
\newcommand{\beq}{\begin{eqnarray}}
\newcommand{\eeq}{\end{eqnarray}}
\newcommand{\mbf}{\mathbf}

\begin{document}
\draft
\tightenlines
\title{General Stability Analysis of Synchronized Dynamics in Coupled Systems}
\author{Yonghong Chen$^{1,3}$,  Govindan Rangarajan$^{2}$
and Mingzhou Ding$^{3}$}

\address{$^1$Institute of Nonlinear Dynamics, Xi'an Jiaotong University,
Xi'an 710049, P.R. China}
\address{$^2$Department of Mathematics and Center for Theoretical
Studies, Indian Institute of Science, Bangalore 560 012, India}
\address{$^3$Center for Complex Systems and Brain Sciences,
Florida Atlantic University, Boca Raton, FL 33431, USA}

\maketitle

\begin{abstract}
We consider the stability of synchronized states (including
equilibrium point, periodic orbit or chaotic attractor) in
arbitrarily coupled dynamical systems (maps or ordinary
differential equations). We develop a general approach, based on
the master stability function and Gershg\"{o}rin disc theory, to
yield constraints on the coupling strengths to ensure the
stability of synchronized dynamics. Systems with specific coupling
schemes are used as examples to illustrate our general method.

\end{abstract}


Large networks of coupled dynamical systems that exhibit
synchronized static, periodic or chaotic dynamics are subjects of
great interest in a variety of fields ranging from
biology\cite{klev} to semiconductor lasers\cite{li1} to
electronic circuits\cite{wu}. For a given problem it is essential
to know the extent to which the coupling strengths can be varied
so that the synchronized state remains stable. Early
attempts\cite{chatt,ding} at this question have typically looked
either at systems of very small size or at very specific coupling
schemes (diffusive coupling, global all to all coupling etc. with
a single coupling strength). Recent work\cite{pecora,hu}
introduced the notion of a master stability function that enables
the analysis of general coupling topologies. This function
defines a region of stability in terms of the eigenvalues of the
coupling matrix. In this paper we present a general method that
provides explicit constraints on the coupling strengths
themselves by combining the master stability function with the
Gershg\"{o}rin disc theory. Our approach is applicable to both
coupled maps and coupled ordinary differential equations (ODEs).
Commonly studied coupling schemes are used as illustrative
examples.

{\it Coupled Maps.} The system we consider is represented by \be
{\mbf x}^i(n+1) = {\mbf f}({\mbf x}^i(n))+\frac{1}{N} \sum_{j=1}^N
G_{ij}\cdot {\mbf H}({\mbf x}^j(n)), \label{gmmap} \ee where
${\mbf x}^i(n)$ is the $M$-dimensional state vector of the $i$th
map at time $n$ and ${\mbf H}: R^M \rightarrow R^M$ is the
coupling function. We define ${\mbf G} = [G_{ij}]$ as the coupling
matrix where $G_{ij}$ gives the coupling strength from map $j$ to
map $i$. The condition $\displaystyle{\sum_j} G_{ij} = 0$ is
imposed to ensure that synchronized dynamics is a solution to
Eq.~(\ref{gmmap}).

Linearizing Eq.(\ref{gmmap}) around the synchronized state ${\mbf
x}(n)$, which evolves according to ${\mbf x}(n+1) = {\mbf f}({\mbf
x}(n))$, we have \be {\mbf z}^i(n+1) = {\mbf J}({\mbf
x}(n))\cdot{\mbf z}^i(n)+\frac{1}{N}\sum_{j=1}^N G_{ij}\cdot
{D\mbf H}({\mbf x}(n))\cdot {\mbf z}^j(n) , \label{gmlinear} \ee
where ${\mbf z}^i(n)$ denotes the $i$th map's deviations from
${\mbf x}(n)$, ${\mbf J}(\cdot)$ is the $M \times M$ Jacobian
matrix for ${\mbf f}$ and $D {\mbf H}(\cdot)$ is the Jacobian of
the coupling function $\mbf H$. In terms of the $M \times N$
matrix ${\mbf S}(n) = ({\mbf z}^1(n) \ {\mbf z}^2(n) \ \cdots \
{\mbf z}^N(n))$, Eq. (\ref{gmlinear}) can be recast as \be {\mbf
S}(n+1) ={\mbf J}({\mbf x}(n))\cdot{\mbf S}(n)+ \frac{1}{N} D
{\mbf H}({\mbf x}(n))\cdot {\mbf S}(n) \cdot {\mbf G}^T,
\label{gmlin2} \ee According to the theory of Jordan canonical
forms, the stability of Eq. (\ref{gmlin2}) is determined by the
eigenvalue $\lambda$ of ${\mbf G}$. Denote the corresponding
eigenvector by $\mbf e$ and let ${\mbf u}(n) = {\mbf S}(n) {\mbf
e}$. Then \be {\mbf u}(n+1) =({\mbf J}({\mbf x}(n))+
\frac{1}{N}\lambda \cdot D {\mbf H}({\mbf x}(n)))\cdot {\mbf
u}(n). \label{gmlinf}\ee So the stability problem originally
formulated in the $M \times N$ space has been reduced to a problem
in a $M \times M$ space where it is often the case that $M \ll N$.
It is worth mentioning that this eigenvalue based analysis is
valid even if the coupling matrix $\mbf G$ is
defective\cite{hirsch}.

We note that $\lambda=0$ is always an eigenvalue of ${\mbf G}$ and
its corresponding eigenvector is $(1\ 1\ \cdots \ 1)^T$ due to the
synchronization constraint $ \sum \limits^N_{j=1}{G_{ij}=0}$. In
this case, Eq.~(\ref{gmlinf}) can be used to generate the Lyapunov
exponents for the individual system, which we denote by
$h_1=h_{max} \geq h_2 \geq \cdots \geq h_M$. These exponents
describe the dynamics within the synchronization manifold defined
by ${\mbf x}^i = {\mbf x} \ \forall i$.

The subspace spanned by the remaining eigenvectors is transversal
to the synchronization manifold, the dynamics in which will be
stable if the transversal Lyapunov exponents are all negative. To
examine this problem, we treat $\lambda$ in Eq.~(\ref{gmlinf}) as
a complex parameter and calculate the maximum Lyapunov exponent
$\mu_{max}$ as a function of $\lambda$. This function is referred
to as the master stability function by Pecora and
Carroll\cite{pecora}. The region in the
$(\rm{Re}(\lambda),\rm{Im}(\lambda))$ plane where $\mu_{max} < 0$
defines a stability zone denoted by $\Omega$. Figure 1 shows a
schematic of two possible configurations of $\Omega$. Whether
$\Omega$ is an unbounded area [Fig.1(a)] or a bounded one
[Fig.1(b)] is contingent on the coupling scheme and other system
parameters. The origin, which is the zero eigenvalue of $\mbf G$,
may or may not lie in the stability zone. For example, for
equilibrium or periodic state in coupled maps, the origin is in
$\Omega$, but for chaos, it lies outside of $\Omega$. We note
that, typically, $\Omega$ is obtained numerically. In some
instances analytical results are possible (see below).

Clearly, if all the transversal eigenvalues of $\mbf G$ lie within
$\Omega$, then the synchronized state is stable. Here we seek
constraints applicable directly to the coupling strengths. This
problem is dealt with by combining the master stability function
with the Gershg\"{o}rin disc theory.

The Gershg\"{o}rin disc theorem\cite{horn} states that all the
eigenvalues of a $n \times n$ matrix ${\mbf A}=[a_{ij}]$ are
located in the union of $n$ discs (called Gershg\"{o}rin discs)
where each disc is given by \be
 \{ z \in C : |z-a_{ii}| < \sum_{j \neq i} |a_{ji}| \}, \ \ \
 i=1,2, \ldots, n.
\ee To apply this theorem to the transversal eigenvalues we need
to remove $\lambda=0$. We appeal to a order reduction technique in
matrix theory\cite{golub} which leads to a $(N-1) \times (N-1)$
matrix $\mbf D$ whose eigenvalues are the same as the eigenvalues
of $\mbf G$ except for $\lambda=0$.

Suppose that, for a given matrix ${\mbf G}$, we have knowledge of
one of its eigenvalues $\tilde{\lambda}$ and the eigenvector $\mbf
e$. Through proper normalization we can make any component of
$\mbf e$ equal one. Here, without loss of generality, we assume
that the first component is made equal 1, namely, ${\mbf
e}=(1,{\mbf e}^T_{N-1})^T$. Rewrite $\mbf G$ in the following
block form: \be
\mbf G=\left(%
\begin{array}{cc}
  G_{11} & {\mbf r}^T \\
  \mbf s & {\mbf G}_{N-1} \\
\end{array}%
\right) \ee with ${\mbf r}=(G_{12},\cdots,G_{1N})^T$, ${\mbf
s}=(G_{21},\cdots,G_{N1})^T$ and \be
{\mbf G}_{N-1}=\left(%
\begin{array}{ccc}
  G_{22} & \cdots & G_{2N} \\
  \vdots & \vdots & \vdots \\
  G_{N2} & \cdots & G_{NN} \\
\end{array}%
\right). \ee
Choose a matrix $\mbf P$ in the form \be
{\mbf P}=\left(%
\begin{array}{cc}
  1 & {\mbf 0}^T \\
  {\mbf e}_{N-1} & {\mbf I}_{N-1} \\
\end{array}%
\right). \ee Here ${\mbf I}_{N-1}$ is the $(N-1) \times (N-1)$
identity matrix. Similarity transformation of $\mbf G$ by $\mbf P$
yields \be
{\mbf P^{-1}G P} =\left(%
\begin{array}{cc}
  \tilde{\lambda} & {\mbf r}^T \\
  {\mbf 0} & {\mbf G}_{N-1}-{\mbf e}_{N-1}{\mbf r}^T \\
\end{array}%
\right). \ee Since $\mbf P^{-1}G P$ and $\mbf G$ have identical
eigenvalue spectra, the $(N-1) \times (N-1)$ matrix \be {\mbf
D}^1={\mbf G}_{N-1}-{\mbf e}_{N-1}{\mbf r}^T \label{gmeq3} \ee
assumes the eigenvalues of $\mbf G$ sans $\tilde{\lambda}$. We can
obtain $N$ different versions of the reduced matrix, which we
denote by ${\mbf D}^k$ $(k=1,2, \cdots ,N)$, depending on which
component of ${\mbf e}$ is made equal 1.

Applying the above technique to the coupling matrix $G$ by letting
$\tilde{\lambda}=0$ and ${\mbf e}=(1\ 1\ \cdots \ 1)^T$ we get
${\mbf D}^k=[d_{ij}^k]$ where $d_{ij}^k=G_{ij}-G_{kj}$. From the
Gershg\"{o}rin theorem the stability conditions of the
synchronized dynamics are expressed as
\begin{enumerate}

\item The center  of every Gershg\"{o}rin disc of ${\mbf D}^k$
lies inside the stability zone $\Omega$. That is,
$(G_{ii}-G_{ki},0) \in \Omega$.

\item The radius of every Gershg\"{o}rin disc of ${\mbf D}^k$ satisfies
the inequality $ \displaystyle{\sum_{j=1, j\neq i}^N\left|G_{ji}-
G_{ki}\right| }< \delta(G_{ii}-G_{ki}),\ \ i = 1, 2, \ldots , N
\quad \mbox{and} \quad i\neq k$. Here $\delta(x)$ is the distance
from point $x$ on the real axis to the boundary of the stability
zone $\Omega$.

\end{enumerate}
As $k$ varies from $1$ to $N$, we obtain $N$ sets of stability
conditions. Each one provides sufficient conditions constraining
the coupling strengths.

{\em Coupled ODE's.} The above procedure for obtaining stability
bounds can also be applied to coupled identical ODEs written as
\beq\label{geneq1} {\mathbf{\dot{x}}}^{i} & = & {\mathbf
F}({\mathbf x}^{i})+\frac{1}{N}\sum_{j=1}^N G_{ij} {\mathbf
H}({\mbf x}^j), \eeq where ${\mbf x}^{i}$ is the $M$-dimensional
vector of the $i$th node. The dynamics of the individual node is
${\mathbf \dot{x}} = {\mathbf F}({\mathbf x})$. Linearizing
around the synchronized state we get \be \dot{{\mbf z}^i} = {\mbf
J}({\mbf x})\cdot {\mbf z}^i + \frac{1}{N} \sum_{j=1}^N
G_{ij}\cdot D {\mbf H}({\mbf x}) \cdot {\mbf z}^j,
\label{linear2}\ee where ${\mbf z}^i$ denotes deviations from
${\mbf x}$, ${\mbf J}(\cdot)$  and $D {\mbf H}(\cdot)$ are the $M
\times M$ Jacobian matrices for the functions of $\mbf F$ and
$\mbf H$. Adopting Jordan canonical form, we obtain \be \dot{{\mbf
u}} = \left[ {\mbf J}({\mbf x}) + \frac{1}{N}\lambda \cdot D {\mbf
H}({\mbf x}) \right] {\mbf u}, \label{linf2}\ee where $\lambda$ is
an eigenvalue of ${\mbf G}$. Performing the same analysis as for
coupled maps, we obtain the same stability conditions as given
above.

{\em Examples.} We now illustrate the general approach by applying
the above results to two examples where analytical results are
possible. In the first example we consider the coupled
differential equation systems with ${\mathbf H}({\mbf x})={\mbf
x}$\cite{hu}. It is easy to see that $D{\mbf H}$ is a $M \times M$
identity matrix. The Lyapunov exponents for Eq. (\ref{linf2}) are
easily calculated since the identity matrix commutes with ${\mbf
J}({\mbf x})$. Denoting them by $\mu_1(\lambda)$,
$\mu_2(\lambda)$, \ldots, $\mu_M(\lambda)$, we have \be
\mu_i(\lambda) = h_i + \frac{1}{N}{\rm {Re}}(\lambda), \ \ \
i=1,2, \ldots ,M. \ee For stability, we require the transversal
Lyapunov exponents ($\lambda \neq 0$) to be negative. This is
equivalent to the statement that the maximum Lyapunov exponent is
less than zero: \be \label{odecond} \mu_{max}(\lambda) = h_{max}
+ \frac{1}{N}\rm {Re}(\lambda) < 0. \ee In other words, the
stability zone $\Omega$ is the region defined by ${\rm
Re}(\lambda) < -Nh_{max}$. The distance function from the center
of each Gershg\"{o}rin disc to the stability boundary is given by:
$\delta(G_{ii}-G_{ki})=-h_{max}-(G_{ii}-G_{ki})$ ($i=1,\ldots,N ,\
\ i\neq k$). Thus the $k$-th set of stability conditions are
\beq (G_{ii}-G_{ki})< -Nh_{max}, \ \ \  \\
\sum_{j=1, j\neq i}^N\left|G_{ji}-G_{ki}\right|
 <-Nh_{max}-(G_{ii}-G_{ki}), \ \ \ \\ i=1,2, \ldots,N ,\ \ i\neq k.
 \nonumber \eeq
It is obvious that the second inequality implies the first one.
So the stability condition for the synchronized state (whether an
equilibrium, periodic or chaotic state) is given by
 \be \label{stblode2}
\sum_{j=1, j\neq i}^N\left|G_{ji}-G_{ki}\right| +(G_{ii}-G_{ki})
< -Nh_{max}, \ \ \ i=1,2, \ldots,N ,\ \ i\neq k. \ee

When the coupling is symmetric, i.e. $G_{ij}=G_{ji}$, Rangarajan
and Ding \cite{raj}, based on the use of Hermitian and positive
semidefinite matrices, derived a very simple stability constraint
\be G_{ij}>h_{max},\quad \forall \ i,j. \label{rajode} \ee We show
here that Eq. (\ref{rajode}) is a consequence of the more general
stability conditions given in Eq. (\ref{stblode2}). This can be
seen as follows. First consider $k=1$. Substituting
$G_{ii}=-\displaystyle{\sum^N_{j=1,j\neq i}} G_{ji}$
(synchronization condition) and simplifying we get the following
equation: \be \sum_{j=2,j\neq i}^N |G_{ji}-G_{1i}|-\sum_{j=2,j\neq
i}^N G_{ji}-2 G_{1i}< -Nh_{max},\ \ \,i\neq 1.  \ee If
$G_{ji}-G_{1i}$ is positive for all allowed $i$ and $j$ values, it
is easy to see that the above stability condition is satisfied
given the condition in Eq. (\ref{rajode}). However, if more than
two such terms are negative we have a problem. We can get around
this by considering the other $(N-1)$ sets of stability conditions
obtained by setting $k=2,3,\ldots,N$ in Eq. (\ref{stblode2}): \beq
\sum_{j=1,j\neq i\neq 2}^N|G_{ji}-G_{2i}|-\sum_{j=1,j\neq i\neq
2}^N G_{ji}-2G_{2i}& < & -Nh_{max}, \ \ \ i\neq 2 \nonumber  \\
 \vdots \; \; \; \; \; \; \; \; \; \; \; \; \; \; \; \\
\sum_{j=1,j\neq i}^{N-1}|G_{ji}-G_{Ni}|-\sum_{j=1,j\neq i}^{N-1}
G_{ji}-2G_{Ni} & < & -Nh_{max}, \ \ \ i\neq N \nonumber \eeq If we
take the average of the inequalities over $k$, cancellation takes
place, resulting in a simplified inequality that will be satisfied
if the sufficient condition given in Eq. (\ref{rajode}) is met. In
other words, the previously derived stability condition is
obtained as a special case when we require the coupling strengths
to meet the $N$ stability conditions simultaneously.

In the second example, we consider a coupled map with $\mbf H=\mbf
f$\cite{ding}. Under this assumption, $D {\mbf H}=\mbf J$ and the
linearized equation [cf. Eq. (\ref{gmlinf})] reduces to \be {\mbf
u}(n+1) =(\lambda/N+1){\mbf J}({\mbf x}(n)){\mbf u}(n).
\label{gmlinfs}\ee The Lyapunov exponents for Eq. (\ref{gmlinfs})
are easily calculated analytically. Denoting them by
$\mu_1(\lambda)$, $\mu_2(\lambda)$, \ldots, $\mu_M(\lambda)$, we
have \be \mu_i(\lambda) = h_i + \ln |\lambda/N+1|, \ \ \ i=1,2,
\ldots ,M. \ee

For stability, we require $\mu_{max}(\lambda) = h_{max} + \ln
|\lambda/N+1| < 0$. In other words, the stability zone is defined
by \be |\lambda+N| < N\exp(-h_{max}) \label{gmstable} \ee The
distance from the center of each Gershg\"{o}rin disc to the
boundary is easily calculated to be
$\delta(G_{ii}-G_{ki})=N\exp(-h_{max})-|N+G_{ii}-G_{ki}|$
($i=1,\ldots,N ,\ \ i\neq k$). Thus the conditions of stability
are
 \beq \label{mapstbl2} \sum_{j=1, j\neq
i}^N\left|G_{ji}-G_{ki}\right| +\left|N+(G_{ii}-G_{ki})\right|<
N\exp(-h_{max}),\ \ \ \\ i=1,\ldots,N ,\ \ i\neq k, \ \ k= 1\ {\rm
or}\ 2 \ {\rm or}\ \cdots\ {\rm or} \ N \nonumber. \eeq For each
$k$ from $1$ to $N$, we obtain a set of sufficient stability
conditions.

In \cite{raj}, a simple stability bound for synchronized chaos in
the case of symmetric coupling was obtained as \be
[1-\exp(-h_{max})]< G_{ij} < [1+\exp(-h_{max})], \quad \forall
i,j. \label{rajmap}\ee This can again be derived from the general
stability condition in Eq.(\ref{mapstbl2}) with the averaging
technique used above.

In summary, we have set up a general formalism to study the
stability of synchronized state in coupled identical maps and
ordinary differential equations. We have also considered the often
used coupling function for coupled maps and coupled ODEs and given
analytical results in these cases. We have also shown that known
stability bounds can be derived from our more general results.

The work was supported by US ONR Grant N00014-99-1-0062. GR was
also supported by the Homi Bhabha Fellowship.

\newpage

\section*{Figure Caption}

\begin{description}
\item{\bf Figure 1:} Schematic illustrations of the stability zone.
(a) unbounded area, (b) bounded area.
\end{description}
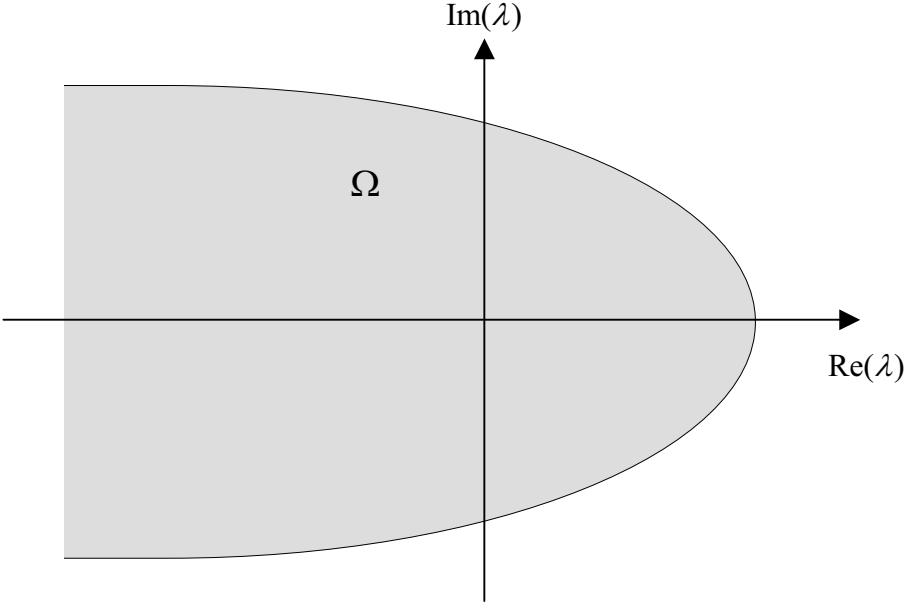
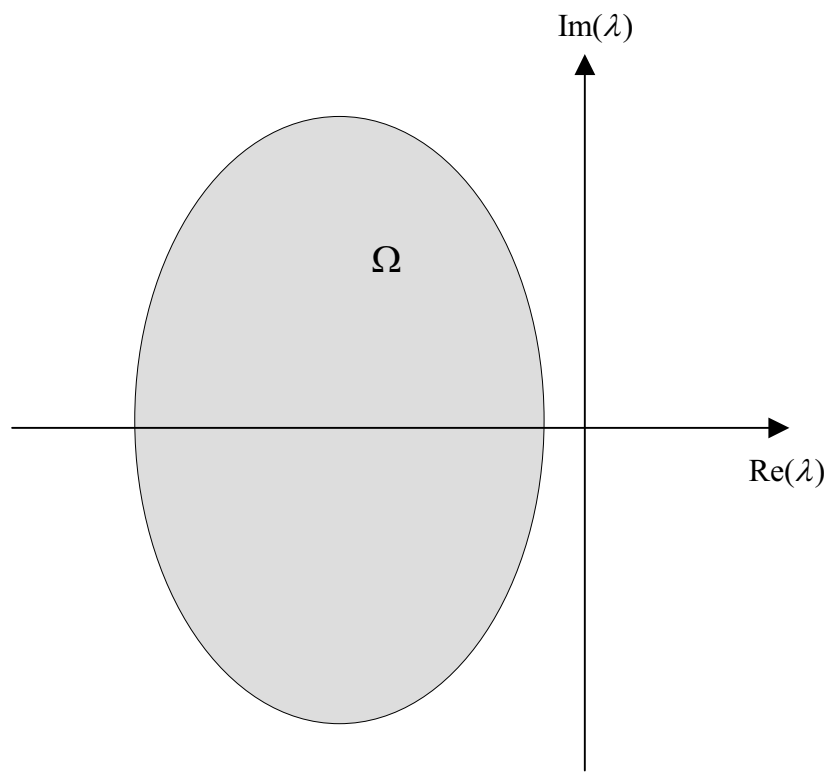
\end{document}